\documentclass[journal,12pt,onecolumn,draftclsnofoot,]{IEEEtran}

\usepackage[cmex10]{amsmath}
\usepackage[utf8]{inputenc}
\usepackage{multirow}
\usepackage{soul}
\usepackage{algorithm}
\usepackage{algpseudocode}
\usepackage{graphicx}
\usepackage{dsfont}
\usepackage{xcolor}
\usepackage{cite}

\usepackage[utf8]{inputenc}
\usepackage{mathtools}
\usepackage[mathscr]{euscript}
\usepackage{amsmath}
\usepackage{amssymb}
\usepackage{amsfonts}
\usepackage{amssymb}
\usepackage{amsmath}
\usepackage{verbatim}
\usepackage[T1]{fontenc}
\usepackage[utf8]{inputenc}
\usepackage{latexsym}
\usepackage{enumerate}
\usepackage{indentfirst}
\usepackage{calligra}
\usepackage{ulem}
\usepackage{graphicx}
\usepackage{array}
\usepackage{float}
\usepackage{bbm}

\usepackage{geometry}

\usepackage{mleftright}
\usepackage{colortbl}
\makeatletter
\let\old@ps@headings\ps@headings
\let\old@ps@IEEEtitlepagestyle\ps@IEEEtitlepagestyle
\def\psccfooter#1{%
 \def\ps@headings{%
 \old@ps@headings%
 \def\@oddfoot{\strut\hfill#1\hfill\strut}%
 \def\@evenfoot{\strut\hfill#1\hfill\strut}%
 }%
 \def\ps@IEEEtitlepagestyle{%
 \old@ps@IEEEtitlepagestyle%
 \def\@oddfoot{\strut\hfill#1\hfill\strut}%
 \def\@evenfoot{\strut\hfill#1\hfill\strut}%
 }%
 \ps@headings%
}
\makeatother

\usepackage{soul}
\usepackage{multirow}
\usepackage{soul,color}
\usepackage{graphicx}
\usepackage{dsfont}
\usepackage{subcaption}

\usepackage[utf8]{inputenc}
\usepackage{mathtools}
\usepackage[mathscr]{euscript}
\usepackage{amsmath}
\usepackage{amssymb}
\usepackage{amsfonts}
\usepackage{verbatim}
\usepackage[T1]{fontenc}
\usepackage[utf8]{inputenc}

\usepackage{latexsym}
\usepackage{enumerate}
\usepackage{indentfirst}
\usepackage{calligra}
\usepackage{ulem}
\usepackage{multirow}

\usepackage{array}
\usepackage{float}
\usepackage{bbm}

\usepackage{eurosym}

\usepackage{hyperref}

\usepackage{caption}
\usepackage{tikz}
\usepackage{pgfplots}

\pgfplotsset{compat=1.8}
\usepgfplotslibrary{statistics}
\pgfmathdeclarefunction{fpumod}{2}{%
 \pgfmathfloatdivide{#1}{#2}%
 \pgfmathfloatint{\pgfmathresult}%
 \pgfmathfloatmultiply{\pgfmathresult}{#2}%
 \pgfmathfloatsubtract{#1}{\pgfmathresult}%
 \pgfmathfloatifapproxequalrel{\pgfmathresult}{#2}{\def\pgfmathresult{5}}{}%
 }

\usepackage{tabularx}
\usepackage{flushend}
\usepackage{textcomp}
\usepackage{xcolor}
\usepackage{import}

\usepackage{listings}
\usepackage{csvsimple}
\usepackage{tabularx}
\usepackage{adjustbox}

\usetikzlibrary{trees,decorations,shadows}
\tikzset{level 1/.style={sibling angle=45,level distance=4mm}}
\usetikzlibrary{arrows.meta}
\usepackage{forest}
\usetikzlibrary{external}

\let\oldtikzexternalgetnextfilename\tikzexternalgetnextfilename \renewcommand{\tikzexternalgetnextfilename}[1]{\oldtikzexternalgetnextfilename{#1}\expandafter\tikzsetnextfilename\expandafter{#1}}

\usepackage{pgfplotstable}
\usepackage[outline]{contour}
\contourlength{1.2pt}

\pgfplotsset{compat=1.13} 

\usetikzlibrary{spy}
\usetikzlibrary{calc}
\usetikzlibrary{fadings}
\usetikzlibrary{patterns}
\usetikzlibrary{shadows}
\usetikzlibrary{mindmap}
\usetikzlibrary{backgrounds}
\usetikzlibrary{shapes.symbols}
\usetikzlibrary{shapes.multipart}
\usetikzlibrary{shapes.geometric}
\usetikzlibrary{automata,positioning}
\usetikzlibrary{decorations.fractals} 
\usetikzlibrary{decorations.markings}
\usetikzlibrary{decorations.pathreplacing}
\usetikzlibrary{decorations.pathmorphing}
\usepackage{svg}
 
\usepackage{bm}

\usepackage{nomencl}
\usepackage{todonotes}
\makenomenclature
\tikzset{edge from parent/.style={segment angle=10,draw}}

\tikzset{
 my rounded corners/.append style={rounded corners=2pt},
}

\renewcommand{\nomgroup}[1]{%
 \ifthenelse{\equal{#1}{O}}{\item[\textit{Operators}]}{%
 \ifthenelse{\equal{#1}{I}}{\item[\textit{Indices}]}{%
 \ifthenelse{\equal{#1}{A}}{\item[\textit{Acronyms}]}{%
 `\ifthenelse{\equal{#1}{V}}{\item[\textit{Variables and parameters}]}{}}}}}
\usepackage{scalerel}
\usetikzlibrary{svg.path}
\definecolor{orcidlogocol}{HTML}{A6CE39}
\tikzset{
 orcidlogo/.pic={
 \fill[orcidlogocol] svg{M256,128c0,70.7-57.3,128-128,128C57.3,256,0,198.7,0,128C0,57.3,57.3,0,128,0C198.7,0,256,57.3,256,128z};
 \fill[white] svg{M86.3,186.2H70.9V79.1h15.4v48.4V186.2z}
 svg{M108.9,79.1h41.6c39.6,0,57,28.3,57,53.6c0,27.5-21.5,53.6-56.8,53.6h-41.8V79.1z M124.3,172.4h24.5c34.9,0,42.9-26.5,42.9-39.7c0-21.5-13.7-39.7-43.7-39.7h-23.7V172.4z}
 svg{M88.7,56.8c0,5.5-4.5,10.1-10.1,10.1c-5.6,0-10.1-4.6-10.1-10.1c0-5.6,4.5-10.1,10.1-10.1C84.2,46.7,88.7,51.3,88.7,56.8z};
 }
}

\newcommand\orcidicon[1]{\href{https://orcid.org/#1}{\mbox{\scalerel*{ \begin{tikzpicture}[yscale=-1,transform shape]
 \pic{orcidlogo};
 \end{tikzpicture}
 }{|}}}}

\begin{document}
\IEEEoverridecommandlockouts
\IEEEpubid{\makebox[\columnwidth]{979-8-3503-9042-1/24/\$31.00 \copyright 2024 IEEE \hfill } 
\hspace{\columnsep}\makebox[\columnwidth]{\hfill }}
%
\title{\huge{
Distribution network reconfiguration for operational objectives: reducing voltage violation incidents and network losses}
}

\author{\IEEEauthorblockN{Geert Mangelschots, Sari Kerckhove, Md~Umar~Hashmi\orcidicon{0000-0002-0193-6703},
and~Dirk~Van~Hertem~\orcidicon{0000-0001-5461-8891}}
 
 \IEEEauthorblockA{\textit{KU Leuven \& EnergyVille},
Genk, Belgium}

 \IEEEauthorblockA{geert.mangelschots@student.kuleuven.be, (sari.kerckhove, mdumar.hashmi, dirk.vanhertem)@kuleuven.be}
 }


\maketitle

\begin{abstract}

As the share of Distributed energy resources (DER) in the low voltage distribution network (DN) is expected to rise, a higher and more variable electric load and generation could stress the DNs, leading to increased congestion and power losses. 
To address these challenges, DSOs will have to invest in strengthening the network infrastructure in the coming decade. This paper looks to minimize the need for flexibility through dynamic DN reconfiguration.
Typically, European DNs predominantly use manual switches. Hence, the network configuration is set for longer periods of time. Therefore, an opportunity is missed to benefit from more short-term dynamic switching.
In this paper, a method is proposed which identifies the best manual switches to replace with remotely controlled switches based on their performance in terms of avoided voltage congestion incidents and DN power losses. The developed method is an exhaustive search algorithm which divides the problem into 3 subsequent parts, i.e. radial configuration identification, multi-period power flow and impact assessment for reconfigurable switch replacement on DN operation. 
A numerical evaluation shows that replacing the two top-ranked switches in the test case reduced the power losses by 4.51\% and the voltage constraint violations by 38.17\%. 
Thus, investing in only a few reconfigurable switches can substantially improve the operational efficiency of DNs.


\end{abstract}

\begin{IEEEkeywords}
Reconfiguration, Reconfigurable switches, manual switches, exhaustive search
\end{IEEEkeywords}

 \pagebreak

\tableofcontents

 \pagebreak

\section{introduction}
With the goal of the European Union to reduce the amount of emission by 55\% in 2030 and reach climate neutrality by 2050 \cite{European_Commission}, big fossil fuel generators will phase out over the coming years and be replaced by renewable energy sources.
Widespread electrification of heating (via heat pumps) and transportation load (via electric vehicles) means that the electricity needs are projected to grow.
This transition will strain the distribution network (DN), which was not designed to withstand this new diversified load and reverse power flow due to distributed generation sources. 
The distribution networks already undergo a lot more congestion and power quality violation incidents \cite{beckstedde2023fit, hashmi2023robust}.


This paper looks at the possibility of providing assistance to the grid for performing operational distribution network reconfiguration (DNR) with reconfigurable switches (RS) or automatic switches. 
The goal of the RS is to improve the operational efficiency of the DNs by reducing the aggregate losses incurred and the number of voltage limit violations.


\subsection{Challenges for distribution networks}\label{intro: LCT}

In previous papers, the problem of distributed energy resources (DER) injection and voltage problems have been studied \cite{damianakis_assessing_2023}. Low voltage (LV) DNs are not able to handle increased penetration installation of renewable energy sources (RES).
They found that most of the current LV DNs aren't yet able to handle a full injection of RES and large-scale integration of DERs.

This paper focuses on PV injection, as this is 
a problem that many DNs face today \cite{beckstedde2023fit}. Given the commitments made in the European Green Deal, the distribution system operators (DSO) will need to invest in the expansion of these grids. Replacing cables with thicker ones is required, but additionally, more active usage of existing grid infrastructure is necessary.

A high PV injection can cause local voltage problems in a DN. There are a couple of solutions which have already been studied, each with its advantages and disadvantages. {Most of these solutions are either too expensive, such as home battery systems proposed in \cite{meuris_managing_2019}, or are wasteful, such as PV curtailment \cite{hashmi2022can}. 
This paper looks into operational DN reconfiguration to decrease the flexibility needs by reducing DN operational issues such as voltage limit violations and line losses of the DN, as quantified in \cite{hashmi2023robust}, and allow integration of more DERs in DNs. }


\subsection{Distribution network reconfiguration}
DNR is a method to change the layout of the distribution network circuit by changing the state of the switches \cite{pfitscher_intelligent_2013}. DNR can be performed with either manual switches or reconfigurable remote switches. 
The manual switches (cannot be controlled remotely) are typically used for seasonal reconfigurations as pointed out in \cite{zidan2013distribution}, however, operational DNR is mostly performed with RS, as these can be controlled either remotely or automatically \cite{liu2019intra}.


For performing operational DNR, there are restrictions and conditions to the configurations, such as radiality and network connectivity. 
In the absence of such constraints being validated, a reliable supply of electricity cannot be guaranteed to all DN consumers.
This means that the method needs to take these into account for performing DNR \cite{pfitscher_intelligent_2013}.


Reconfiguration has already been implemented into the high voltage \cite{amin_optimal_2023} and medium voltage transmission grid \cite{mendoza_microgenetic_2009}. 
In this work, RS is used for performing DNR to improve operational efficiency for low-voltage DNs by reducing the number of voltage violation incidents and DN losses.


\pagebreak

\subsection{Problem statement}

The DNR problem tackled in this paper is based on the classic reconfiguration problem with multiple switches in a power system that allows topology changes. The main cases considered are static, dynamic and hybrid versions. These are all quite similar and have been solved using a mixed integer formulation \cite{jabr_minimum_2012} or a metaheuristic one \cite{gerez_behavior_2021},\cite{mendoza_microgenetic_2009}.
This paper aims to quantify the implication of replacing static switches with reconfigurable switches on the improved operation of DNs for aggregate DN losses and voltage violations.




\subsection{Contributions}
The contributions of the paper are:


$\bullet$\textit{ Proposed DNR framework}: 
In this paper, a new, computationally efficient, framework for an exhaustive search in reconfiguration is introduced.  The proposed DNR calculation framework decouples the large reconfiguration problem into 3 smaller sub-problems that can be solved sequentially. This limits the search space for each of them.
These sub-problems are: (i) eliminating non-radial configurations, (ii) simulating the circuit with power flows, and (iii) comparing combinations of RS placement.

$\bullet$\textit{ Numerical evaluation}:
 The proposed DNR framework is demonstrated in a test case with realistic load profiles for residential and commercial consumers.

This paper is organized as follows:
Section \ref{sec:radial_conf} details the identification of all possible radial configurations.
Section \ref{section3} details the power flow calculations.
Section \ref{sec:replacement} describes the replacement framework for RSs.
Section \ref{sec:KPI} details the key performance indicators used for evaluating the numerical results detailed in Section \ref{sec:test_case}.
Section \ref{section6} concludes the paper.

\pagebreak

\section{identify radial configurations}
\label{sec:radial_conf}

DNs are typically operated radially. That being said, DNs are designed in a manner that the topology can be adjusted through opening and closing connections, e.g. in an open ring configuration.
A radial distribution network features a tree-like structure where each load has a unique path to the substation, simplifying the protection system and operations.
Graphically, the DN includes a substation, a main feeder line, branches to individual loads, and may incorporate transformers and switches.

In the first part of the proposed method, all non-radial configurations are eliminated through a hierarchical approach. This is done in two steps. Firstly, a reduction of the network graph. Secondly, the radiality conditions are checked.

\subsection{Reducing the network to a graph}

\label{sec:reducing_network}


First, the algorithm converts the DN to a reduced-order graph. This increases the computation speed of the radiality step. The reduction allows only looking at the relevant information without losing any configurations while ensuring algorithmic efficiency. This is done in 3 steps:

\subsubsection*{Step 1} 
The algorithm connects all the feeders to one super node. This is necessary because the transformers are used as a reference bus in the simulation, and connecting these would make the system non-solvable. This also helps the algorithms to find loops as the algorithm now has a stable starting point, i.e. the super node.

\subsubsection*{Step 2}
Next, it reduces the DN by representing all the buses, connected with a branch without a switch, as one graph node. It is assumed that the nodes which are reduced down are all connected radially, as the system otherwise can't be radial \footnote{\underline{Removing switches from the algorithm:}
the algorithm also considers whether a switch is {a switch} between a graph node and itself. In this case, the switch will be opened and ignored to make radial configurations. 
Another way the algorithm limits the number of radial configurations is by looking at multiple switches in a loop. If there are multiple switches in a loop, then these switches are put in connection with each other and one of them always needs to be open.}. 

\subsubsection*{Step 3}

The last step is replacing the branches with switches on them into an edge, yielding a reduced graph.

\subsection{Finding the radial configurations}\label{radial_configurations}

\begin{figure}
    \centering
    \includegraphics[width = 0.7\textwidth]{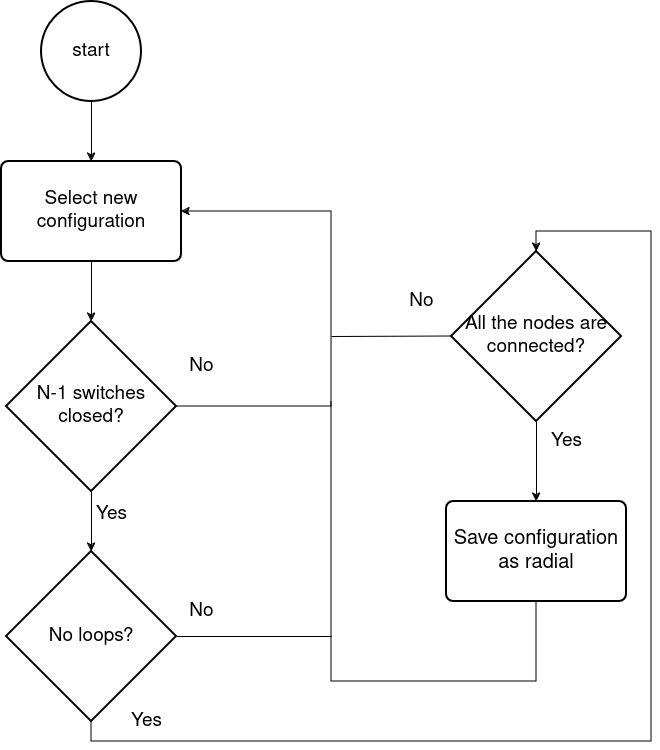}
    \caption{The flowchart of the algorithm to check for radiality}
    \label{fig:radial_flowchart}
\end{figure}

After the system is simplified as described in \ref{sec:reducing_network}, the algorithm represented in figure \ref{fig:radial_flowchart} identifies all of the radial configurations. There are $2^n$ possible configurations where n is the number of switches.
In figure \ref{fig:radial_flowchart}, first the number of closed switches is checked. Then a search algorithm checks whether there are loops in the system. And finally, with the solution of the loop-finding protocol, the algorithm checks that every node is connected to the super node.




\subsubsection{N-1 switches closed condition}

The number of switches is checked to be equal to the number of graph nodes minus 1.


\subsubsection{No loops condition}
A search algorithm finds the available paths from the start node to each node. If more than one path is found than there is a loop in the system. 

\subsubsection{Bus Connection Check}
For the previously identified configuration, the list of all the connected nodes is checked against the list of all the nodes. If these contain the same nodes, everything is connected.

\pagebreak

\section{Power flow \& operational network states }
\label{section3}

The selected radial configurations need to be ranked based on their operational merits. 
In order to assess this, multi-period
power flows are calculated for all radial configurations. The network states identified via the power flows are used to qualitatively rank the configurations for different operational objectives. 

\subsection{Power flow}

After the radial configurations are identified, the power flows are calculated for each of the radial configurations over each time step of the load profiles. These power flows are calculated using the PowerModels.jl package\cite{coffrin2018powermodels}.

These power flows are all simulated using the PowerModels.jl library \cite{coffrin2018powermodels}.
\begin{equation}\label{eq:PF}
    S_i^* =  \tilde{V_i^*} \cdot \sum^n_{j = 1}\tilde{y_{ij}}\cdot \tilde{V_{j}}
\end{equation}

To find the power flow, it uses equations which are derived from equation \ref{eq:PF}. In equation \ref{eq:PF} $\tilde{V_i^*} $ and $ \tilde{V_j}$ are the voltages on bus i and bus j with $\tilde{y_{ij}}$ being the admittance of the cable between these buses. 

Subsequently, the power flow output 
are used to find
the best combination of switches that need to be converted to reconfigurable ones.

\subsection{Objectives}
\label{objectives}
Power loss reduction and voltage violation reduction are set as the objectives, as these give a good overview of how a system can help with the integration of DER \cite{damianakis_assessing_2023}.
\subsubsection{Power losses}

Equation \ref{eq: power_losses_obj} calculates the power loss. In this formula, X is the list of all the radial configurations and x is an element of this list. $N_g$ is the list of all the buses with generators connected to them in the network. $N_l$ is the list of all the buses with loads connected to them in the network and $P_g$ is the generation at each node with $P_l$ being the load at each node. 
\begin{equation} \label{eq: power_losses_obj}
    \min_{\forall x \in X} obj(x) = \sum_i^{N_{g}} P_g(i) - \sum_i^{N_{l}} P_l(i)
\end{equation}

\subsubsection{Voltage constraint violations}
A voltage violation instance is considered if the nodal voltage goes outside the permissible range of 0.95 to 1.05 per unit value.
If this is the case, an additional voltage constraint violation is counted.

\pagebreak

\section{Possible reconfiguration cases}
\label{sec:replacement}

As the algorithm just ran all of the possible power flows, it can calculate the objective value for each time step and configuration. With these objective values, it can start to evaluate the different replacements of switches with the objectives obtained from every case. This will be done by first finding the start configuration. Secondly, it will define all the replacement cases possible. With these results, it will find all reachable configurations for each case. It uses these configurations to find the optimal path and evaluates the cases on the objective values of these paths.


\subsection{Finding the start configuration}
It is crucial to select the start configuration as this affects the model convergence time. 
To find the start configuration, a static reconfiguration of the whole period is done. Subsequently, we pick the configuration with the \underline{lowest average power losses} as the starting configuration.


\subsection{List of replacement cases}

\label{sec:cases}
A replacement case is a combination of static switches replaced by reconfigurable ones. These do not include any combinations with only 1 reconfigurable switch, as this doesn't allow reconfiguration. When the system has more switches, the number of combinations will increase exponentially. These are enumerated as this makes it possible to evaluate each case with the others.

\subsection{Finding reachable configurations}

For each case, the reachable configurations are determined by matching the state of the static switches from the start configuration with the reachable configurations. 

\subsection{Finding optimized paths }

To compare the different cases, the optimal path for each scenario must be determined. This involves independently optimizing each time step by evaluating the objective value for every configuration and comparing them to identify the one with the lowest objective value.


Based on the identified objective for each configuration and time step, the optimized path is calculated. To find the optimized path through the objectives, the path-finding algorithm can look at each time step independently. 

\subsection{Selecting best cases}

For each case created in section \ref{sec:cases} the optimal path is found. This comes with a total objective for the whole period. These objectives are compared to the cases with the same amount of possible reconfigurable switches and the best one is selected. The objectives of the best cases are compared to each other to see which case has the highest impact on the objectives and will thus have the highest gain per RS.  

\section{Numerical evaluation }

\label{section5}

\subsection{Key performance Indicators}
\label{sec:KPI}

For the evaluation of the numerical results, 2 KPI's were chosen:  

\begin{itemize}
    \item Computation time: split into the run time of each subproblem: discussed in section \ref{sec:radial_conf},\ref{section3} and \ref{sec:replacement}
    \item Objective values: reduction of power losses and voltage constraint violations chosen in section \ref{objectives}.
\end{itemize}

\subsection{Test case circuit}
\label{sec:test_case}
\begin{figure}
    \centering
    \includegraphics[width = 0.92\linewidth]{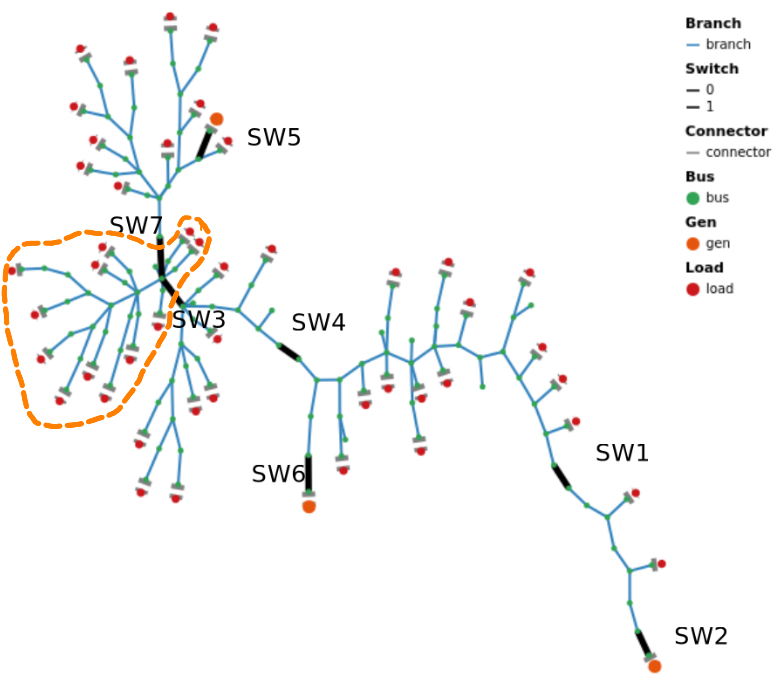}
    \caption{The circuit for the test case\cite{koirala_non-synthetic_2020}.}
    \label{fig:grid_test_case_1}
\end{figure}

This test case is a modification of a sub-circuit of a Spanish circuit \cite{koirala_non-synthetic_2020} as represented in figure \ref{fig:grid_test_case_1}. This system is made up of the following components.  

\begin{itemize}
    \item 3 generators
    \item 7 switches 
    \item 1 commercial district (encircled by the orange dashed line)
    \item 4 residential districts (These are the other areas divided by switches)
\end{itemize}

The diversity of the feeders and the number of switches make this a very interesting case for reconfiguration. By adding the commercial district, the total load will shift a lot during the day.


\subsection{Scenario generation}

\subsubsection{Load profiles}

\begin{figure}
    \centering
    \includegraphics[width = 0.95\linewidth]{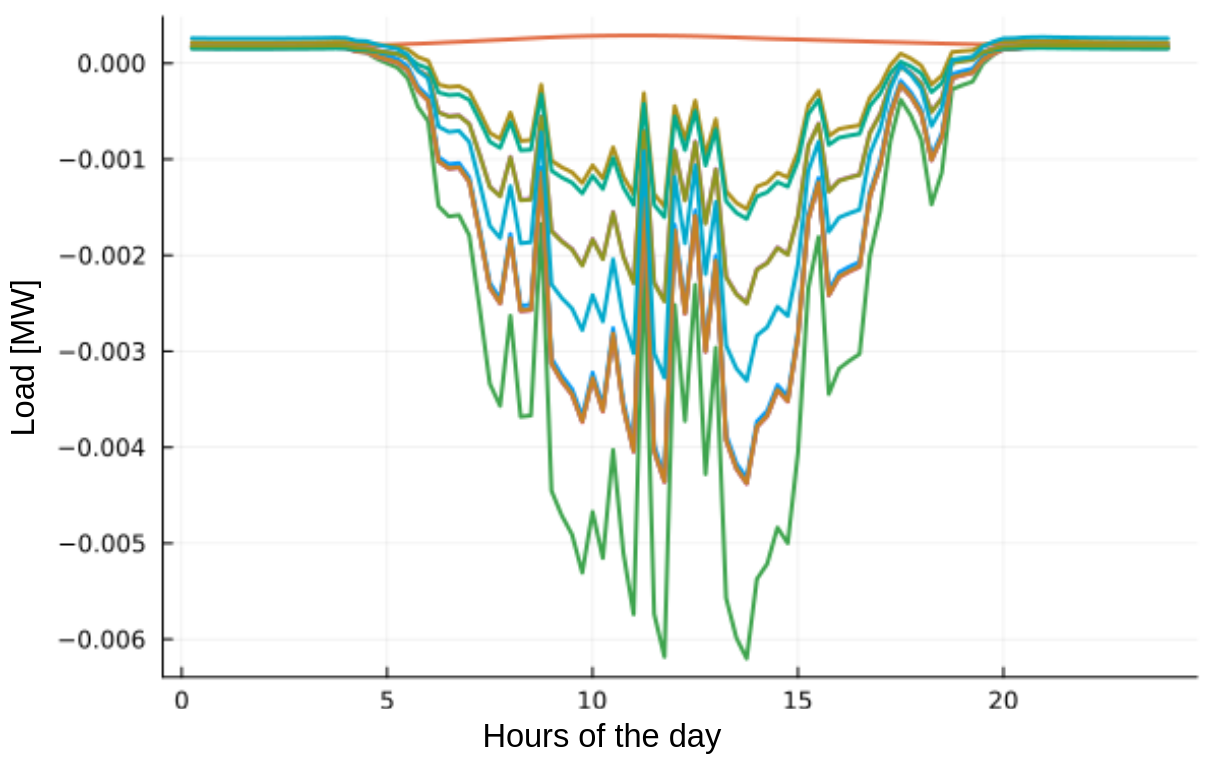}
    \caption{Commercial load profiles}
    \label{fig:commercial_LP}
\end{figure}

\begin{figure}
    \centering
    \includegraphics[width = 0.95\linewidth]{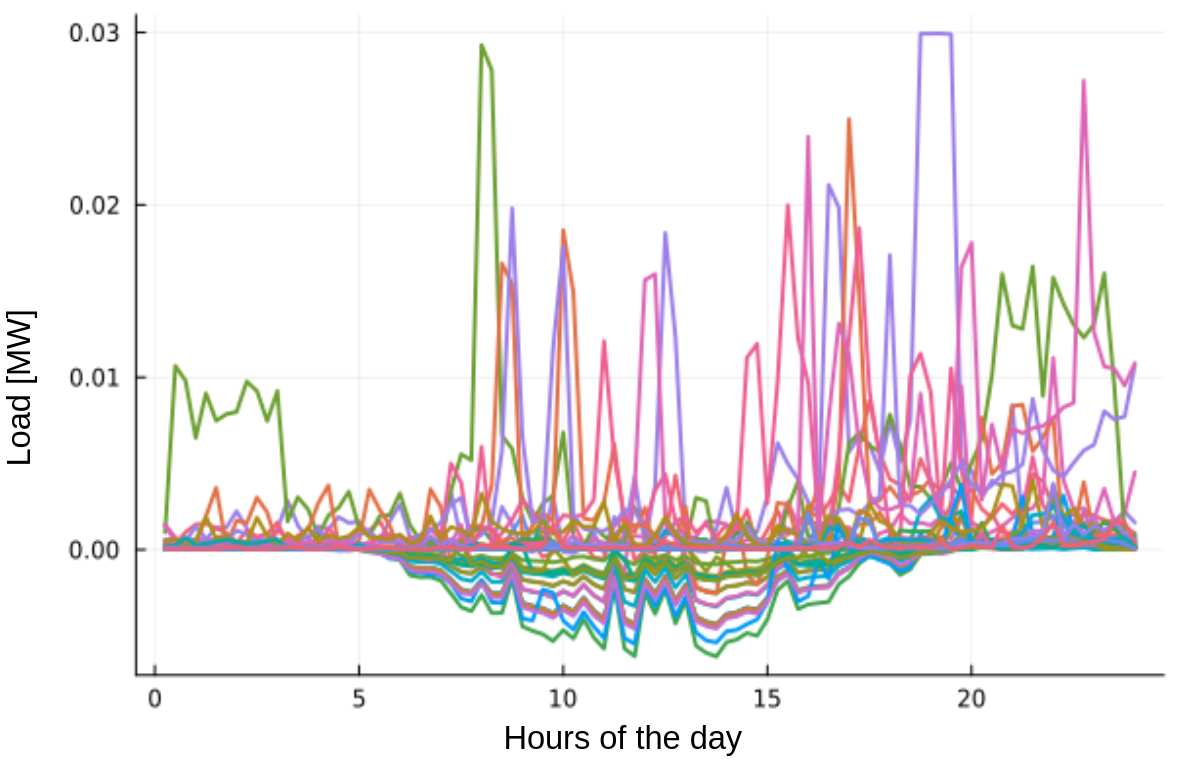}
    \caption{Residential load profiles}
    \label{fig:residential_LP}
\end{figure}

This test case has 2 different types of load profiles. 
The residential load profiles shown in Fig. \ref{fig:commercial_LP} is taken from the
ENWL dataset \cite{navarro-espinosa_dissemination_2015} and the commercial load profile shown in figure \ref{fig:residential_LP} is taken from the OEDI dataset \cite{US_department_of_energy}. 

\textit{Commercial load profiles}:
Figure \ref{fig:commercial_LP} shows that the commercial district has a very high PV injection and will even produce power during the day. This is done to give a future perspective on how high PV injections will influence the power systems' behaviour. This PV data comes from the Elia dataset \cite{hashmi2024analyzing}. Also, the load profiles without any PV injection in them, show that these loads shift only slowly and have a really smooth progression. 

\textit{Residential load profiles}:
Figure \ref{fig:residential_LP} shows that the load profiles from residential buildings are much more variable and have higher peaks. This also shows that more peaks happen in the evening when people are normally at home.

\subsubsection{Adding of switches}

There were 2 switches added to the original Spanish circuit: switches 3 and 7. 

\subsection{Results}

This section will review the intermediate and final results of the test case.

\subsubsection{Radiality results}

The first part of the algorithm finds all of the radial configurations as mentioned in section \ref{sec:radial_conf}. For this example, there are 128 configurations out of which 14 are radial. This means that 89\% of all the configurations are discarded.

\subsubsection{Power flow results}

\begin{figure}
    \centering
    \includegraphics[width = 0.95\linewidth]{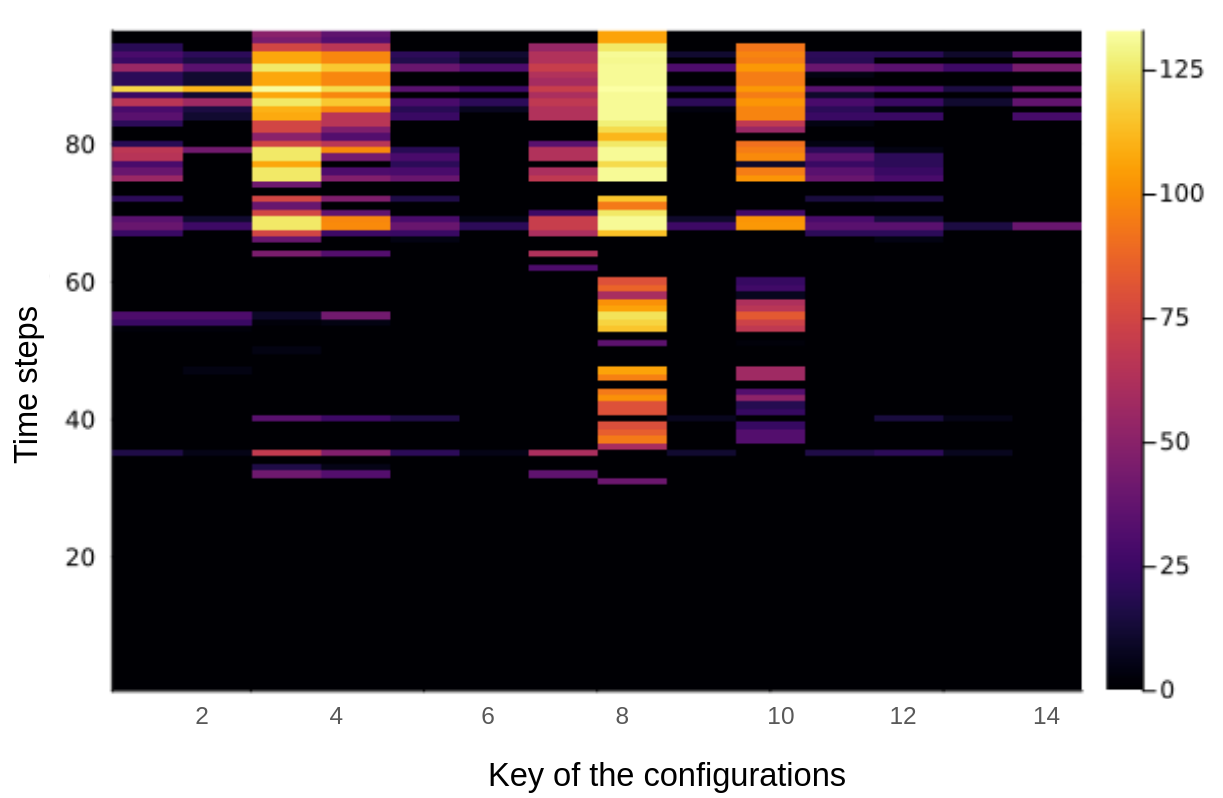}
    \caption{The heatmap of the voltage constraint violations for 14 configurations over 96 time steps }
    \label{fig:heatmap_voltage}
\end{figure}

\begin{figure}
    \centering
    \includegraphics[width = 0.95\linewidth]{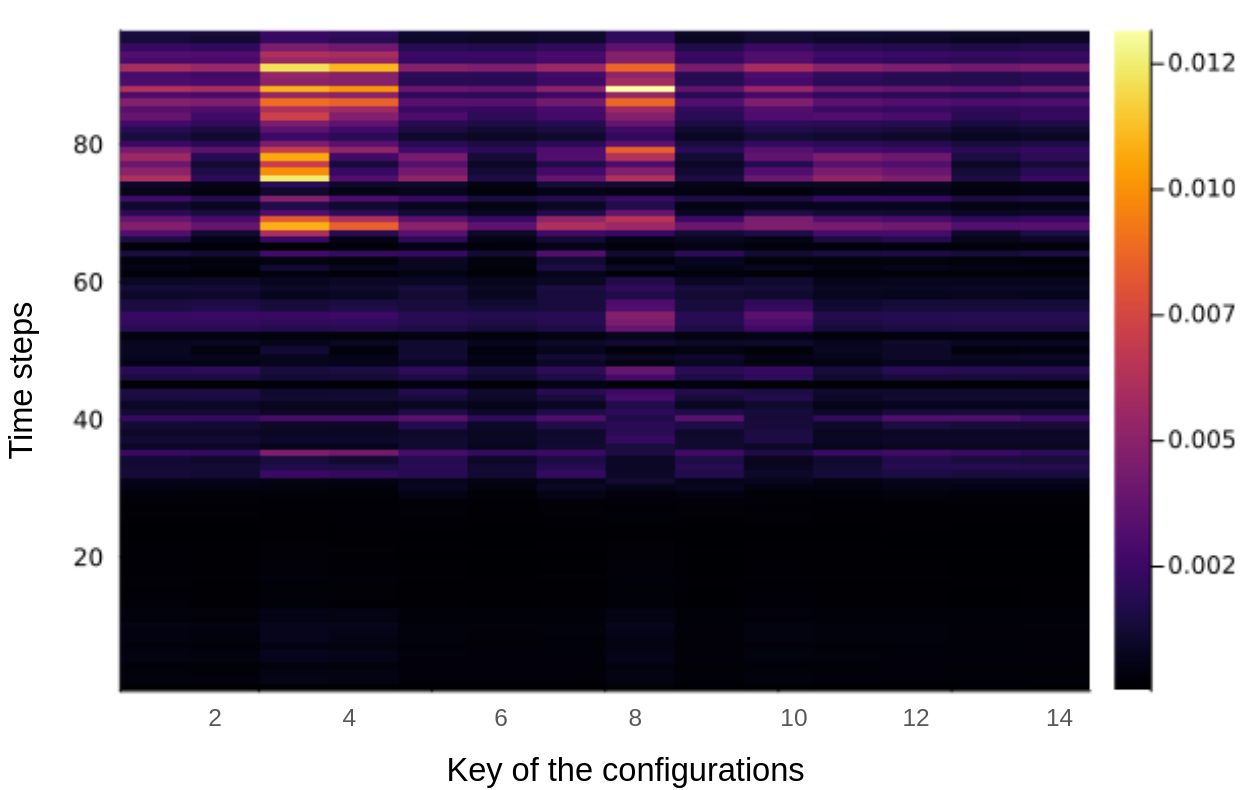}
    \caption{The heatmap of the power losses in MW for 14 configurations over 96 time steps }
    \label{fig:heatmap_power_loss}
\end{figure}

After the algorithm runs the power flows, it retains the power losses and voltage constraint violations for each radial configuration at each time step. The heatmaps in figures \ref{fig:heatmap_voltage} and \ref{fig:heatmap_power_loss} represent these. 


These heatmaps demonstrate that some configurations have consistently worse results. They also show that reconfiguration can not solve every voltage constraint violation but can reduce them drastically.

\subsubsection{Final results}

The last results given from the model are related to the installed RS and the reduction in power losses and voltage constraint violations.

\pagebreak

\textbf{Power losses}

\begin{figure}
    \centering
    \includegraphics[width = 0.9\linewidth]{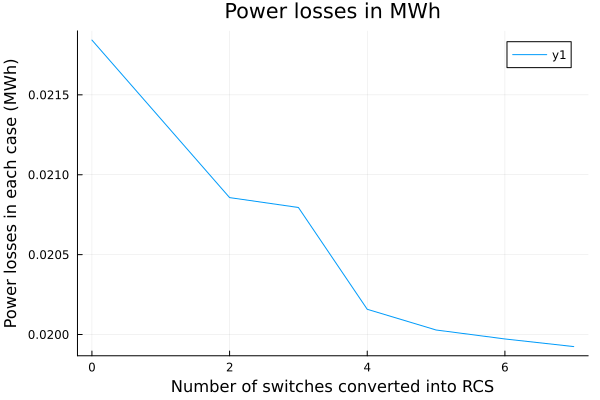}
    \caption{The power losses accumulated during the whole day as a function of the amount of RS installed.}
    \label{fig:results_power_losses_without_CSW}
\end{figure}

Table \ref{tab:replacing_power_losses} indicates which cases are the best for power loss reduction for each number of reconfigurable switches installed. The accumulated power losses in MWh of these cases are represented in figure \ref{fig:results_power_losses_without_CSW}.  It is clear that the biggest reduction in losses happens when the first 2 switches are installed. By replacing switch 7 and switch 4, the commercial district can connect to either one or the other feeder.  

The first 2 switches make a reduction of 4.51\% possible. By adding 5 switches and making all the switches reconfigurable, a reduction of 8.78\% is observed. These percentages were calculated with equation \ref{eq:loss_reduction_1}. In this equation $P_{L_{\text{static}}}$ are the losses in the static case and $P_{L_{\text{dynamic}}}$ are the losses in the dynamic case.
\begin{equation}\label{eq:loss_reduction_1}
    \text{Dynamic}_{\text{loss reduction}} = \frac{ P_{L_{\text{static}}} - P_{L_{\text{dynamic}}}}{P_{L_{\text{static}}}} = 8.78\%
\end{equation}

\begin{table}[!htbp]
    \centering
    \caption{\small{Merit order for switch replacement for reducing power losses.}}
    \begin{tabular}{p{12mm}|c|c|c|c|c|c|c}
        \hline
        \# switches to replace  & SW 1  & SW 2 & SW 3 & SW 4 & SW 5 & SW 6 & SW 7 \\
        \hline
        \hline
        2 & 0 & 0 & 0 & 1 & 0 & 0 & 1\\
        \hline
        3 & 0 & 0 & 0 & 1 & 0 & 1 & 1\\
        \hline
        4 & 1 & 0 & 0 & 1 & 0 & 1 & 1\\
        \hline
        5 & 1 & 1 & 0 & 1 & 0 & 1 & 1\\
        \hline
        6 & 1 & 1 & 1 & 1 & 0 & 1 & 1\\
        \hline
    \end{tabular}
    \label{tab:replacing_power_losses}
\end{table}

\textbf{Voltage constraint violations}

Table \ref{tab:replacing_voltage} shows which switches would optimally be replaced, for each number of switches considered. 

\begin{figure}
    \centering
    \includegraphics[width = 0.9\linewidth]{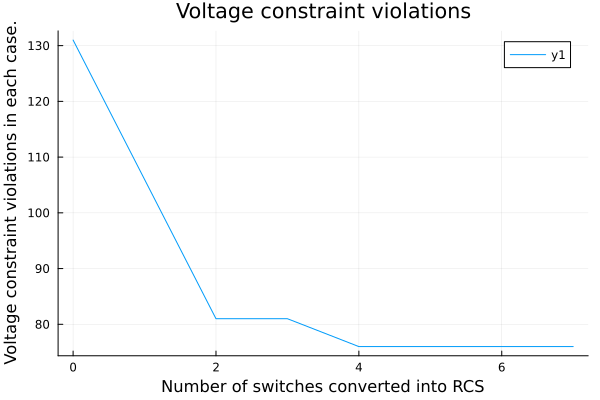}
    \caption{The voltage constraint violations during the whole day as a function of the amount of RS installed without switch flip cost.}
    \label{fig:results_voltage_constraint_violation_without_CSW}
\end{figure}

Figure \ref{fig:results_voltage_constraint_violation_without_CSW} again shows that the biggest decrease of voltage constraint violations can be achieved when replacing just 2 switches. The decrease in voltage constraint violations is 38.17\% when 2 switches are installed and 41.98\% when 4 switches are installed. These percentages were calculated with equation \ref{eq:VV_reduc}. In these calculations $N_{V_{static}}$ are the voltage violations in the static case and $N_{V_{dynamic}}$ are the voltage violations in the dynamic case.
\begin{equation}\label{eq:VV_reduc}
    \text{Dynamic}_{VVreduct} = \frac{N_{V_{static}}-N_{V_{dynamic}}}{N_{V_{static}}} = 41.98\%
\end{equation}

In some cases, replacing an extra switch does not reduce the voltage constraint violations. This can be explained by how the objective is defined. The objective only looks at voltage deviations of more than 5\%.  This means that even though replacing 3 or even 7 switches will enable an increase in the quality of the voltage profile, it might not decrease the number of voltage constraint violations.


\begin{table}[!htbp]
    \centering
    \caption{\small{Merit order for switch replacement for reducing voltage constraint violations.}}
    \begin{tabular}{p{12mm}|c|c|c|c|c|c|c}
        \hline
        \# switches to replace  & SW 1  & SW 2 & SW 3 & SW 4 & SW 5 & SW 6 & SW 7 \\
        \hline
        \hline
        2 & 0 & 0 & 0 & 1 & 0 & 0 & 1\\
        \hline
        4 & 1 & 0 & 0 & 1 & 0 & 1& 1\\
        \hline
    \end{tabular}
    \label{tab:replacing_voltage}
\end{table}

A comparison of tables \ref{tab:replacing_power_losses} and \ref{tab:replacing_voltage} makes it apparent that the same switches need to be replaced for the reduction of both power losses and voltage constraint violations in the case of 2 switches being replaced. 

To make sure that there are no unsolvable problems, after running the calculation, a check is done to ensure the system can be switched out of extreme voltage constraint violations. In this test case, a deviation of 10\% from 1 p.u. is seen as an extreme voltage constraint violation. This was not the case in this test case.

\subsection{Evaluation}

This model will be evaluated by referring back to the KPI's in section \ref{sec:KPI}. This section proposed 2 performance indicators to evaluate the numerical results, elaborated next.

\subsubsection{Computation time}


To find an accurate computation time, a large number of Monte Carlo simulations are executed. For this analysis run times of 100 simulations for this test case and a simple system were used. 
These timing results are divided into the 2 main parts of each simulation: finding the radial configurations and running all power flows. The average time of these parts is listed in table \ref{tab:timing_results}. To investigate scaling ability, a simple synthetic test case was used to compare the run times. This table indicates that the most influential part of the computation time will be the calculation of the power flow. The power flow calculations are however already reduced as they are only calculated for the radial configurations. The radiality protocol eliminates almost 90\% of the total number of configurations on which otherwise power flows needed to be calculated. The power flow subroutine is the only part of the problem which scales badly. This part however is very parallelizable and a better scaling can be tried using threads running on different cores.

\begin{table}[!htbp]
    \centering
    \caption{\small{Average computation time per part in each test case over 100 simulations.}}
    \begin{tabular}{p{7mm}|p{12mm}|p{10mm}|p{13mm}|p{10mm}|p{10mm}}
    \hline
        test case & \# of radial conf. & \# buses & \# switches& radial  (s) & power flow(s)\\
        \hline \hline
        Simple case & 5 & 6 & 4&3.3 & 4.8  \\ 
        \hline
        Spanish case & 14 & 138 &7 & 3.4 & 100.2 \\ 
        \hline
    \end{tabular}
    \label{tab:timing_results}
\end{table}
\subsubsection{Objective values}

The values in figure \ref{fig:results_voltage_constraint_violation_without_CSW} show that the voltage violations have been reduced by 41.98\% when all of the switches are replaced and by 38.17 \% when only 2 switches are replaced. Figure \ref{fig:results_power_losses_without_CSW} shows a reduction of 4.51\% when 2 switches are replaced and 8.78\% when all the switches are replaced. This means that the voltage congestion objective is the most important reduction in this case study.

\pagebreak

\section{Conclusion}
\label{section6}

Given the increased stress on the distribution network (DN), the transition to operational reconfiguration is assessed as an option to reduce power losses and voltage constraint violations. Typically, current DN mostly have manual switches which can be configured for long periods of time. These manual switches do not enable the possible benefits of short-term switching.
This paper compares the possible reductions per number of RS installations, enabling DSO's to plan topological changes to the network. 

The proposed hierarchical exhaustive search-based method for operational network reconfiguration splits up the problem into 3 subsequent subroutines: (i) reducing the search space by eliminating all non-radial configurations, (ii) simulating the circuit with power flows, and (iii) comparing combinations of reconfigurable switch placement. Despite exhaustively evaluating the power flow of each radial configuration for each time step, the method results in an acceptable computation time and is scalable considering the parallelization of the power flow. The numerical evaluation over a test case shows that introducing 2 remote switches results in a potential reduction of the total power losses equal to 4.51\% and a reduction of the voltage constraint violations of 38.17\%. The replacement of all 7 switches yielded a total loss reduction of 8.78\% and reduced the voltage constraint violations by 41.98\%. Thus, a small number of reconfigurable switches substantially impacts DN operations.

In future work, we will consider a cost-benefit analysis considering the cost of switch flips, power loss and voltage constraint violations.


\pagebreak

\section*{Acknowledgement}
This work is supported by the Flemish Government and Flanders Innovation \& Entrepreneurship (VLAIO) through the \href{https://www.improcap.eu/}{IMPROcap} project (HBC.2022.0733).


\pagebreak

\bibliographystyle{IEEEtran}
\bibliography{reference}

\end{document}